\documentstyle[12pt,epsfig,amstex,amsfonts,amssymb]{article}
\newcommand{\bm}[1]{\mbox{\boldmath $#1$}}
\title{On the {\em strong} anomalous diffusion}
\author{P.~Castiglione$^{1}$,
 A.~Mazzino$^{2,3}$, P.~Muratore-Ginanneschi$^{4}$ and 
A.~Vulpiani$^{1}$\\
\small $^{1}$ 
 Dipartimento di Fisica, and Istituto Nazionale di Fisica della Materia,\\
\small Universit\`a ``La Sapienza'', P.le A. Moro 2, 00185
Roma, Italy.\\
\small{$^{2}$ CNRS, Observatoire de Nice, B.P. 4229,
06304 Nice Cedex 4, France.}\\
\small$^3$ INFM - Dipartimento di Fisica, Universit\`a di Genova, I--16146 
Genova, Italy\\
\small$^{4}$
N.B.I., Blegdamsvej 17 DK 2100, Copenhagen \O, Denmark
}

\begin{document}
\maketitle
\date{}

\begin{abstract}
        The superdiffusion behavior, i.e. $<x^2(t)> \sim t^{2 \nu}$, 
        with $\nu > 1/2$, in general is not completely characherized by 
        a unique exponent. We study  some systems exhibiting strong anomalous
        diffusion,
        i.e. $<|x(t)|^q> \sim t^{q \nu(q)}$ where $\nu(2)>1/2$ and 
        $q \nu(q)$ is not a linear function of $q$.
        This feature is different from the weak superdiffusion regime, i.e.   
        $\nu(q)=const > 1/2$, as in  random shear flows.
      
        The strong anomalous diffusion can be generated by
        nontrivial chaotic dynamics, e.g. Lagrangian motion in $2d$
        time-dependent incompressible velocity fields, $2d$ symplectic
        maps and $1d$ intermittent maps. Typically the function
        $q \nu(q)$ is piecewise linear. This corresponds to two
        mechanisms: a weak anomalous diffusion for the typical events
        and a ballistic transport for the rare excursions.

        In order to have  strong anomalous diffusion one needs a violation
        of the hypothesis of the central limit theorem, this happens
        only in a very narrow region of the control parameters space.

        In the presence of the strong anomalous diffusion one
        does not have a unique exponent and therefore one has the
        failure of the usual scaling of the probability
        distribution, i.e. $P(x,t)=t^{-\nu}F(x/t^{\nu})$.
        This implies that  the effective  equation  at 
        large scale and long time for  $P(x,t)$, cannot obey 
        neither the  usual Fick equation nor other linear equations
        involving  temporal and/or spatial fractional derivatives.

\end{abstract}
PACS number(s): 05.45.+b, 05.60.+w; \\

\section{Introduction}

The transport of mass or heat from particles passively advected 
by a given velocity field is a problem of considerable practical
and theoretical interest.
The coupling between advection and molecular diffusitivy results in
a great variety of different fields, as geophysics, chemical engineering
and  disordered media \cite{Moffa,BouchGeo}.
One can have highly nontrivial behavior
even in the presence of simple  laminar velocity fields,
 e.g. very large diffusion coefficients due to the combined effects
of the molecular diffusitivy and some features of the velocity fields
\cite{Moffa,Crisall}.

Taking into account the molecular diffusion, the Lagrangian motion of a 
test particle is described by the Langevin equation :
\begin{equation}
\dot{\bm x}={\bm u}({\bm x},t)+{\bm \xi} 
\label{11}
\end{equation}
where ${\bm u}({\bm x},t)$ is the Eulerian velocity field at the position 
${\bm x}$ and the time $t$, and ${\bm \xi}$ is a Gaussian white noise with 
zero mean and correlation function 
\begin{equation}
\langle \xi_i(t) \xi_j(t') \rangle= 2 D_0 \delta_{ij} \; \delta(t-t'),  
\label{12}
\end{equation}
the coefficient $D_0$ being the molecular diffusivity.
If $\Theta$ is the density of tracers, the Fokker--Planck 
equation associated to (\ref{11}) is
\begin{equation}
\partial_t \Theta +{\bm \partial} ({\bm u}\;\Theta ) = 
D_0 \partial^2 \Theta.  
\label{13}
\end{equation}
For times much larger than the typical time of ${\bm u}$, the large-scale
density field $\langle\theta\rangle$ (i.e. the field $\theta$ averaged
over a volume of linear dimension much larger than the typical length
of the velocity field ${\bm u}$) obeys a standard diffusion Fick equation:
\begin{equation}
\partial_t \langle \Theta \rangle = D^E_{ij} \partial^2_{x_i x_j} 
\langle \Theta \rangle \;\;\;\;\;\; i,j=1,\ldots,d.  
\label{14}
\end{equation}
All the (often nontrivial) effects due to the presence of the velocity field 
are in the eddy diffusion coefficient $D^E_{ij}$.
Of course if the Eq.~(\ref{14}) holds then one has
 $\langle (x(t)-x(0))^2 \rangle \simeq 2 \; D^E_{11}\; t$. 
In practice at large time the test particle behaves as a Brownian particle.

The above scenario is the typical one.
Nevertheless there exist cases where anomalous diffusion is observed, i.e.
 $\langle(x(t)-x(0))^2\rangle \sim  t^{2 \nu}$ with $\nu \neq 1/2$.
The case when $\nu < 1/2$ corresponds to subdiffusion while for  
$\nu > 1/2$ one has superdiffusion. 
Trapping is the basic mechanism leading to subdiffusion and it occurs  
only in compressible fields.
A remarkable example of  subdiffusion,  $\langle (x(t)-x(0))^2 \rangle 
\sim  [\ln(t)]^4 $
is the random walk  in $1d$ random potential \cite{Sinai}.

If the velocity field is incompressible and the molecular diffusivity is non 
zero, either standard diffusion or superdiffusion takes place.

In the last two decades many authors have studied systems with
anomalous (non Gaussian) diffusion.  It is practically impossible to
cite the huge literature.  We just mention, among the many, the
contribution of Geisel and coworkers \cite{Geiseall} for the $1d$
intermittent maps, Ishizaki et al \cite{Ishi} for the symplectic maps,
Zaslavky et al \cite{Zasla} for the lagragian motion in q-flows
\cite{MatdeMar,Zumo}, Majda \& Avellaneda \cite{AveMai} for the random
shear flows, the experimental study of Solomon at al. \cite{SoloWeek}
on the superdiffusion in an anular tank, the symbolic dynamics
approach of Misguich et al. \cite{Misgui} for the subdiffusive
behavior in a stochastic layer and the anomalous diffusion of magnetic
lines in the $3d$ turbulence by Zimbardo et al. \cite{Zimba}.

The aim of this paper is a detailed study of the anomalous diffusion
in simple velocity fields and discrete time dynamical systems. 
In particular we shall show that {\em strong} anomalous diffusion 
may occur, i.e.
\begin{equation}
\langle (x(t)-x(0))^q \rangle \sim  t^{q \,\nu(q)}
\label{15}
\end{equation}
where $\nu(q)$ is non-constant. In this case the probability
distribution $P(x,t)$ cannot be described in term of a unique
scaling exponent, as in the L\'evy flight \cite{Motroll} or in the 
random shear flow.

Section 2 is devoted to a brief review of the known mechanisms for
anomalous diffusion.  In Sec.~3 we report detailed numerical results
for the anomalous diffusion mechanism in $2d$ time-dependent
incompressible velocity field, in the standard map (which can be
considered a discrete time version of the Lagrangian motion in a $2d$
time periodic incompressible velocity field), and in an intermittent
$1d$ map.  In Sec.~4 the reader can find some conjectures and
discussions.

\section{A brief review on some aspects of the anomalous diffusion}

\label{s:review}
Anomalous diffusion occurs when some, or all, of the hypothesis of the 
central limit theorem break down.
Practically, the system has to violate at least one of the two following 
conditions :
\begin{itemize}
\item[a)] finite variance of the velocity;
\item[b)] fast enough decay of the auto-correlation function of 
Lagrangian velocities.
\end{itemize}
This is, of course, an elementary remark.
Nevertheless, in our opinion, a detailed discussion of this point 
can be useful.
In the present paper we use the term anomalous diffusion to indicate a non 
standard 
diffusion in the asymptotic limit of very long time. Sometimes in the 
literature the term anomalous is used also for long (but non asymptotic) 
transient behavior.
This transient anomalous behavior can have practical relevance, e.g. in 
geophysics. We do not discuss this point in details.
Let us recall some well established results on the relation between diffusion 
phenomena and correlation function.
It is easy to obtain the following formula :
\begin{equation}
\langle \left( x_i(t)-x_i(0)\right)^2 \rangle = \int_0^t dt_1 \int_0^t dt_2 
\langle v_i({\bm x}(t_1)) \; v_i({\bm x}(t_2))\rangle \simeq 2\; t\; \int_0^t 
d\tau \; C_{ii} (\tau)
\label{21}
\end{equation}
where $C_{ij}(\tau)$ is the correlation function of the Lagrangian velocity,
namely
${\bm v}=\dot{\bm x}$,
\begin{equation}
C_{ij}(\tau)=\langle \; v_i ({\bm x}(\tau))\; v_j ({\bm x}(0)) \; \rangle.
\label{22}
\end{equation}
Eq.~(\ref{21}) is the so-called Taylor relation \cite{Taylor21} 
for a test particle evolving 
according to the Langevin equation (\ref{11}).

For test particles whose dynamics is described by
a map (discrete time),
\begin{equation}
{\bm x}(n+1)={\bm F}({\bm x}(n)),
\label{24}
\end{equation}
 the Taylor formula (\ref{21}) is still valid with the obvious changes :
${\bm v}({\bm x}(t)) \mapsto {\bm F}({\bm x}(n)) -{\bm x}(n)$ and 
$\int_0^t C_{ii}(\tau) \mapsto C_{ii}(0)/2+\sum_{j=1}^n C_{ii}(j)$.
If both $\langle v^2\rangle < \infty$ and $\int_0^t d\tau \; C_{ii}(\tau) 
< \infty$
then one has standard diffusion and the effective diffusion coefficients 
are 
\begin{equation}
D^E_{ii}=\lim_{t \rightarrow \infty} \frac{1}{2 \; t} \langle \left( x_i(t) - 
x_i(0) \right)^2 \rangle = \int_0^{\infty} \; d \tau \; C_{ii}(\tau)
\label{25}
\end{equation}
 From the above remarks it stems that anomalous diffusion occurs only in two 
circumstances: 
\begin{itemize}
\item[a)] $\langle v^2 \rangle = \infty$
\item[b)] $\langle v^2 \rangle < \infty$ and $\int_0^t d \tau \;
 C_{ii}(\tau) = \infty $, i.e. $C_{ii}(\tau) \sim \tau^{-\beta}$ with  
$\beta > 1$, that means very strong correlations.
\end{itemize}  

\subsection{Simple systems exhibiting anomalous diffusion}

The L\'evy flight \cite{Motroll} belongs to the first case.
Let us briefly discuss it for discrete one dimensional systems. 
The position $x(t+1)$ at the time $t+1$ is obtained from $x(t)$ as follows: 
\begin{equation}
x(t+1)=x(t)+U(t),
\label{26}
\end{equation} 
where $U(t)$'s are independent variables identically distributed according 
to a  $\alpha$--L\'evy--stable distribution, 
$P_{\alpha}(U)$, with the following well-known properties :
\begin{equation}
\int \; d U e^{i k U} P_{\alpha}(U) \propto e^{-c \mid k \mid^{\alpha}} 
\;\;\;\;
{\rm and} \;\;\;\; P_{\alpha}(U) \sim  U^{-(1+\alpha)} \;\;\;\; {\rm for} 
\;\;\;\; \mid U \mid \gg 1  
\label{211b}
\end{equation}
with $0 < \alpha \leq 2$.
An easy computation gives 
\begin{equation}
\langle x(t)^q \rangle = \left\{
\begin{array}{cc}
C_q\;t^{q/\alpha}      & q < \alpha \\   
\infty      &  q \geq \alpha.
\end{array}
\right.
\label{27}
\end{equation}
Note that $\langle x^2 \rangle = \infty$ for any $\alpha < 2$, nevertheless 
one can consider the L\'evy flight as a sort of anomalous diffusion in the sense 
that $x \sim t^{1/\alpha} \gg t^{1/2}$.

In spite of the relevance of the $\alpha$--L\'evy--stable distribution  
in probability theory, the L\'evy flight model, in our opinion, has a rather 
weak importance for physical systems. This is due to the very unrealistic 
property of infinite variance.

An elegant way to overcome the above trouble is the introduction of a 
stochastic model, called L\'evy walk \cite{Schlei}, 
corresponding to (\ref{26}) but now 
$U(t)$ is a random variable with nontrivial correlations. The velocity 
$U(t)$ can  assume the values $\pm u_0$ and it maintains its value for a 
duration $T$ which is a random variable with probability density $\psi(T)$.   
Practically, the particle moves with constant velocity
for a certain time  after which  
it changes direction and so on. In the L\'evy walk the origin of the 
possible anomaly is transferred to the correlation function of the Lagrangian 
velocity.
Actually, taking
\begin{equation}
\psi(T)\sim T^{-(\alpha+1)},
\label{28}
\end{equation}
one has standard diffusion if $\alpha > 2$ while for $\alpha < 2$ one has 
anomalous (super) diffusion :
\begin{equation}
\langle x(t)^2 \rangle \sim t^{2 \nu} \;\;\;\;\; \nu=\left\{
\begin{array}{cc}
1/2             &  \alpha > 2  \\
(3-\alpha)/2    & 1 < \alpha < 2 \\   
1               & \alpha< 1.
\end{array}
\right.
\label{29}
\end{equation}
Sometimes in the literature the L\'evy walk is (erroneously) called L\'evy 
flight.

\subsection{Nontrivial anomalous diffusion}

 From the above considerations one realizes that it is not particularly 
difficult to build up {\em ad hoc} probabilistic models exhibiting 
anomalous diffusion.
On the contrary it is much more difficult and much more interesting the 
understanding of the anomalous diffusion in nontrivial systems like the 
transport 
in incompressible velocity fields or deterministic maps, where the explicit 
probabilistic aspect (given for example by an external noise related to 
the molecular diffusivity) cannot play a relevant role.

Avellaneda and Majda \cite{AveMajda}, (see also Avellaneda \& Vergassola 
\cite{Averga} for the 
generalization to the time dependent case), obtained a very important 
and general result about the diffusion in an incompressible velocity field 
${\bm u}({\bm x})$.
If the molecular diffusivity $D_0$ is non zero and the 
infrared contribution 
to the velocity field are weak enough, namely 
\begin{equation}
\int d{\bm k} \; \frac{\langle \mid \hat{{\bm u}}({\bm k}) \mid^2 \rangle}
{k^2} < \infty
\label{210}
\end{equation}
where $\langle \; \cdot \; \rangle$ indicates the time average and 
$\hat{{\bm u}}$ 
represents the Fourier transform of the velocity field, 
then one has standard diffusion, i.e. the diffusion coefficients $D_{ii}$'s in 
(\ref{25}) are finite.
Therefore there exist two possible origins for the superdiffusion 
\begin{itemize}
\item[a)] $D_0 > 0$ and, in order to violate the (\ref{210}),
the velocity field with very long spatial 
correlation;
\item[b)] $D_0=0$ and strong correlation between ${\bm u}({\bm x}(t))$ and 
${\bm u}({\bm x}(t+\tau))$ at large $\tau$.
\end{itemize}
For the sake of completeness we briefly remind one of the few nontrivial 
systems for which 
the presence of anomalous diffusion can be proven in a rigorous way.

Consider a $2d$ random shear flow :
\begin{equation}
{\bm u}=\left( u(y)\;,\; 0 \right)
\label{211}
\end{equation} 
where $u(y)$ is a random function such that 
\begin{equation}
u(y)=\int_{-\infty}^{\infty} d k \; e^{i k y} \; \hat{u}(k) \;\;\;\;\;\;\; 
\langle \hat{u}(k) \; \hat{u}(k')\rangle = S(k) \; \delta(k-k'),
\label{212}
\end{equation}
$S(k)$ is the spectrum and the average $\langle \; \cdot \; \rangle$ is taken 
over the field realizations.
Matheron \& De Marsily \cite{MatdeMar} showed  that the anomalous diffusion 
in the 
$x$-direction occurs if
$$ \int d k \; \frac{S(k)}{k^2} =\infty.$$
On the contrary, if this integral is finite one has standard diffusion and
\begin{equation}
D^E_{11}=D_0 +\frac{1}{D_0} \int_0^{\infty} dk \; \frac{S(k)}{k^2}.
\label{213}
\end{equation}
Eq.~(\ref{213}) is a well-known exact result already obtained by Zeldovich 
\cite{Zeldo}. 
For the sake of simplicity we consider the case 
\begin{equation}
S(k)\sim k^{\gamma} \;\;\; \;\; k \mapsto 0.
\label{214}
\end{equation}
If $\gamma > 1$ then one has standard diffusion.
On the contrary, if $ -1 \leq \gamma \leq 1$ one has a superdiffusion, namely
\begin{equation}
\langle \; \mid x(t)-x(0) \mid^2\; \rangle \sim t^{2 \nu} \;\;\;\;\;\; 
\nu = \frac{3-\gamma}{4} \geq \frac{1}{2}.
\label{215}
\end{equation}
The condition $\int d k \; S(k)\;k^{-2} =\infty$ for the anomalous
diffusion can be understood by means of the following simple physical
argument. Note that $\int d k \; S(k)\;k^{-2}\sim \langle u^2\rangle
\;L^2$ where $L$ is the typical length of the function $u(y)$,
i.e. the typical distance between two sequent zeros of $u(y)$. If $
\langle u^2\rangle < \infty$ and $\int d k \; S(k)\;k^{-2} < \infty$
the diffusion process is basically similar to that one characterized
by a velocity field given by a sequence of strips of size $L$ and
velocity $\pm \sqrt{\langle u^2\rangle }$ and therefore the
(\ref{213}) is nothing but the transversal Taylor diffusion
\cite{Taylor54} in channels.  The origin of the anomalous diffusion in
this case is due to the fact that a test particle travels in a given
direction for a very long time before it changes direction and so on.
In some sense the Lagrangian motion in the random shear flow for $-1
\leq \gamma \leq 1$ is a nontrivial realization of a L\'evy walk.  The
generalization to the $3d$ case ${\bm u}=\left( u(y,z),0,0 \right)$ is
straightforward.

Another nontrivial system where the presence of anomalous diffusion
has been rigorously proven is the Kraichnan's  passive scalar model 
\cite{K94}, where  the velocity  advecting the scalar field, 
is rapidly varying in time. For such a model, 
generally, the distance between pairs of
particles tends to increase with the time elapsed but, occasionally,
particles may come very close and stay so; this is the source of
the anomalies in the scaling \cite{BGK97} \cite{FMV98}.

%As far as we know the random shear flows are the unique nontrivial systems 
%with continuous time evolution, for which the anomalous diffusion has been 
%rigorously proven.

If the inequality (\ref{210}) holds then in order to have anomalous 
diffusion one 
needs $D_0=0$ and a correlation function $C_{ii}(\tau)$ going to zero not 
too fast. These conditions are satisfied, e.g., in time periodic velocity 
fields whose Lagrangian phase space has a complicated self-similar structure 
of island and cantori \cite{Zasla}. 
In this case superdiffusion is essentially 
due to the almost trapping of the trajectories for arbitrarily long time, 
close to the cantori which are organized in complicated self-similar 
structures.
   
In addition, there exist a strong numerical evidence (and sometimes also 
theoretical arguments) for superdiffusion in $1$-d intermittent maps as 
well as in the standard map, which is a $2$-d symplectic map and therefore 
it can be considered as the Poincar\'e map of a Lagrangian motion in a 
time periodic velocity field. 

\subsection{About the meaning of {\em anomalous}}

Up to now we used the term anomalous as synonymous of non-Gaussian. 
Let us consider now in more detail the cases with superdiffusion.
 There are two possibilities :
\begin{itemize}
\item[a)]weak anomalous diffusion when 

\begin{equation}
\langle \mid x(t)-x(0) \mid^q \rangle \sim t^{q \nu} 
\;\;\;\;  \forall q > 0 \;\; {\rm and} \;\; \nu>\frac{1}{2} 
\label{216}
\end{equation}
\item[b)]strong anomalous diffusion when
\begin{equation}
\langle \mid x(t)-x(0) \mid^q  \rangle \sim t^{q \,\nu(q)} \;\;\;\; 
\nu(q) \neq cost \;\; \; \nu(2) > \frac{1}{2} 
\label{217}
\end{equation}
where $\nu(q)$ is a nondecreasing function of $q$.
\end{itemize}
In the weak anomalous diffusion one has for the probability distribution 
$P(\Delta x,t)$ of 
$\Delta x =x(t)-x(0)$ at time $t$,  the following expression :
\begin{equation}
P(\Delta x,t)= t^{-\nu} F(\Delta x \; t^{-\nu})
\label{218}
\end{equation}
where in general the function $F$  is different from the Gaussian one.
The random shear flow is an example of the weak anomalous diffusion
\cite{Zumo}.

Bouchaud at al. \cite{Boucall} have shown that in the case of the random shear flow (\ref{211})-(\ref{213}) with $\gamma=0$ 
\begin{equation}
F(a) \sim e^{-c \mid a \mid^{4/3} } \;\;\; {\rm for} \;\; \mid a\mid \gg 1.
\label{219}
\end{equation} 

There are very few numerical studies on the strong anomalous diffusion.
Let us mention the work of Sneppen \& Jensen \cite{SneppJe} on the motion of 
particles passively advected by the dynamical membranes (the authors 
use the term multidiffusion) and an $1$-d intermittent map studied by Pikovsky 
\cite{Pikov}.

Of course the strong anomalous diffusion is not compatible with the simple 
scaling (\ref{218}).
Obviously in the presence of anomalous diffusion $P(x,t)$ cannot obey to the 
usual Fickian diffusion Eq.~(\ref{24}). Therefore, a quite natural problem
 is to find the effective evolution equation for $P(x,t)$ at large $t$.
The answer to this question up to now as far as we know, is not well 
understood in the general case.

For the L\'evy flight one can naively write 
\begin{equation}
\partial_t P=-c\; (-\partial^2)^{\alpha/2} \; P.
\label{225}
\end{equation}  
The above equation can be considered the proper one for the L\'evy flight in 
the sense that starting at $t=0$ with $P(x,0)=\delta(x)$ at any $t>0$ 
the probability $P(x,t)$ is given by th $\alpha$-stable L\'evy function.
Of course, if $\alpha=2$ one has the usual Fick equation and the Gaussian 
shape for $P(x,t)$.
For the more complicated anomalous diffusion there are many proposal in terms 
of fractional time, and/or spatial, derivatives. We do not enter in a detailed 
discussion of these works, the interested readers can see \cite{Russi}. 

\section{Numerical results}
\label{s:numres}
In this section we report detailed numerical results for anomalous diffusion 
in $2$-d time dependent velocity fields and maps.

\subsection{{\em Strong} anomalous diffusion in a 
simple flow mimicking the Rayleigh--B\'enard convection}
We investigate here anomalous diffusion in a simple 
model mimicking the Rayleigh--B\'enard convection \cite{GS88a,GS88b}.
Two-dimensional convection with rigid boundary conditions is described by 
the following stream function:
\begin{equation}
\psi (x,y,t) = \psi_0 \sin (x + B\sin\omega t ) \sin y \;\;\; ,
\label{gollub}
\end{equation}
where the periodicity of the cell is $L\equiv 2\pi$, and the
even oscillatory instability
\cite{CB74} is accounted for by the term $B\sin \omega t$, representing
the lateral oscillation of the rolls.
The capability of the simple flow (\ref{gollub}) to capture the essential
features of the convection problem is discussed in Ref.~\cite{GS88b}.

It is worth noting that, at fixed $B$, the only relevant parameter controlling 
the diffusion process in the flow (\ref{gollub})
is $\epsilon \equiv\omega L^2 /\psi_0$, i.e.,
the ratio between 
the lateral roll oscillation frequency  
and the characteristic frequency of the passive scalar motion.
Different regimes
take place for different values of $\epsilon$. The two limiting cases 
$\epsilon \ll 1$ and $\epsilon\gg 1$ have been investigated 
in Ref.~\cite{CNRY91}. The regime 
$\epsilon\sim 1$ has recently been studied in Ref.~\cite{CCMVV98},
in particular the synchronization between the lateral roll oscillation 
frequency (of the order of $\omega$) and the characteristic frequencies 
of the scalar field motion (of the order of $\psi_0/L^2$) has been 
investigated.
This mechanism leads to enhanced diffusion.

We briefly recall the main results of Ref.~\cite{CCMVV98}. 
It has been found that the synchronization
between the circulation in the cells and their global oscillation
is a very efficient way of jumping from cell to
cell. This mechanism, similar to stochastic resonance
\cite{BSV81,BPSV83}, makes the effective diffusivity
as a function of the frequency $\omega$ very
structured. In the limit where the molecular diffusion vanishes,
anomalous superdiffusion takes place for narrow windows of values
of $\omega$ around the peaks.

\begin{figure}
\begin{center}
\mbox{\psfig{file=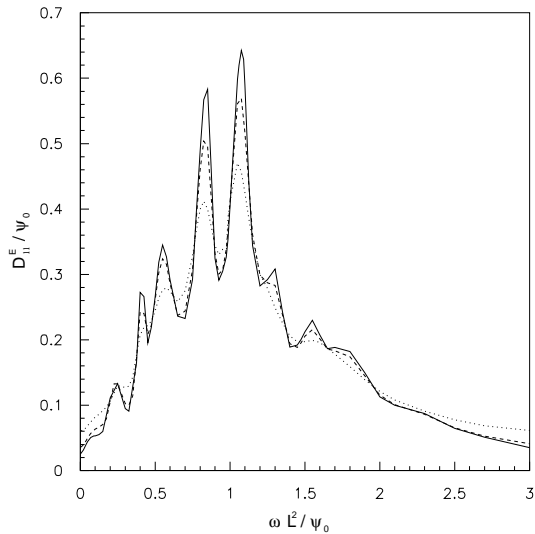,height=7cm,width=7cm}}
\end{center}
\caption{The turbulent diffusivity $D^E_{11}/\psi_0$ {\it vs} 
the frequency $\omega L^2 /\psi_0$ for different values of the 
molecular diffusivity $D_0/\psi_0$. 
$D_0/\psi_0=3\times 10^{-3} $ (dotted curve);
$D_0/\psi_0=1\times 10^{-3} $(broken curve); 
$D_0/\psi_0=5\times 10^{-4} $ (full curve).}
\label{calenda}
\end{figure}

In Fig. \ref{calenda} the $x$-component, $D^E_{11}$, of the eddy diffusivity 
versus the frequency $\omega L^2/\psi_0$ is shown for different values 
of $D_0/\psi_0$.

We anticipate that anomalous diffusion
in the {\em strong} sense (according to the definition given in Sec.~2.2)
takes place, making highly nontrivial
the diffusion process only around the values of $\omega$ corresponding to the 
peaks of enhanced diffusion. Specifically, 
let us focus our attention on the anomalous regime occurring 
for $D_0=0$ and $\omega L^2/\psi_0\simeq 1.1$. We have then integrated the 
Eq.~(11) with $D_0=0$ and ${\bm u}$ obtained from the flow 
(\ref{gollub}), using a second-order Runge-Kutta scheme. 
In the following, averages are over different realizations and performed by 
uniformly distributing $10^6$ particles in the basic periodic cell. 
The system evolution is computed up to times of the order of 
$10^4\;L^2/\psi_0$. 

In order to investigate the anomalous
diffusion in the strong sense, a measure of 
$\langle   [x(t) - x(0) ]^q\rangle$ for different $q$'s has to be made.
 
%------------------------------------------------
\begin{figure}
\begin{center}
\mbox{\psfig{file=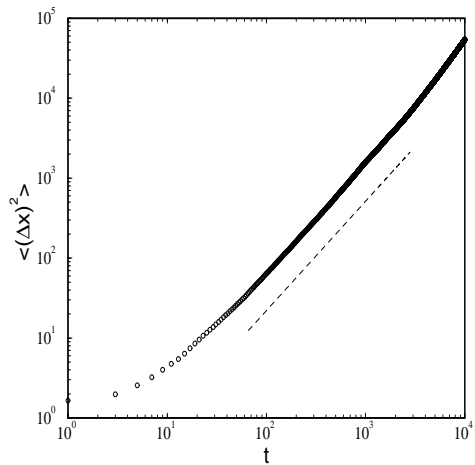,height=7cm,width=7cm}}
\end{center}
\caption{The mean-squared displacement {\em vs} the time  
for the flow
(\protect\ref{gollub}) with $D_0=0$, 
and $\omega = 1.1$. 
Lengths 
and times are shown in units of $L$ and $ L^2/\psi_0$, respectively.
The best-fit (dashed) line corresponds to $2 \nu(2) = 1.3$.}
\label{d_vs_t}
\end{figure}
%------------------------------------------------
To that end,  we have performed 
a linear fit of $\ln \langle   [x(t) - x(0) ]^q\rangle$ 
{\em versus} $\ln t$, the slope of which
being the exponent $q\,\nu(q)$ in (22). As an example, in 
Fig.~\ref{d_vs_t} the mean-squared
displacement, as a function of the time, has been presented. 
Superdiffusive transport takes thus place for $\omega=1.1$ 
in the limit $D_0\to 0$,
i.e.:
\begin{equation}
\langle   [x(t) - x(0) ]^2\rangle \propto t^{2 \nu(2)} \qquad \mbox{with}
\qquad \nu(2) > 1/2 \;\;\; . 
\label{superlinear}
\end{equation}
The presence of genuine anomalous diffusion is also supported by the
behavior of the coefficient $D^E_{11}$ as a function of  $D_0$. 
As suggested in Ref.~\cite{BCVV95}, in the presence of
genuine anomalous diffusion, the effective diffusivity must
diverge and it is expected that $D^E_{11} \sim D_0^{-\beta}$ with $\beta
> 0$.
%------------------------------------------------
\begin{figure}
\begin{center}
\mbox{\psfig{file=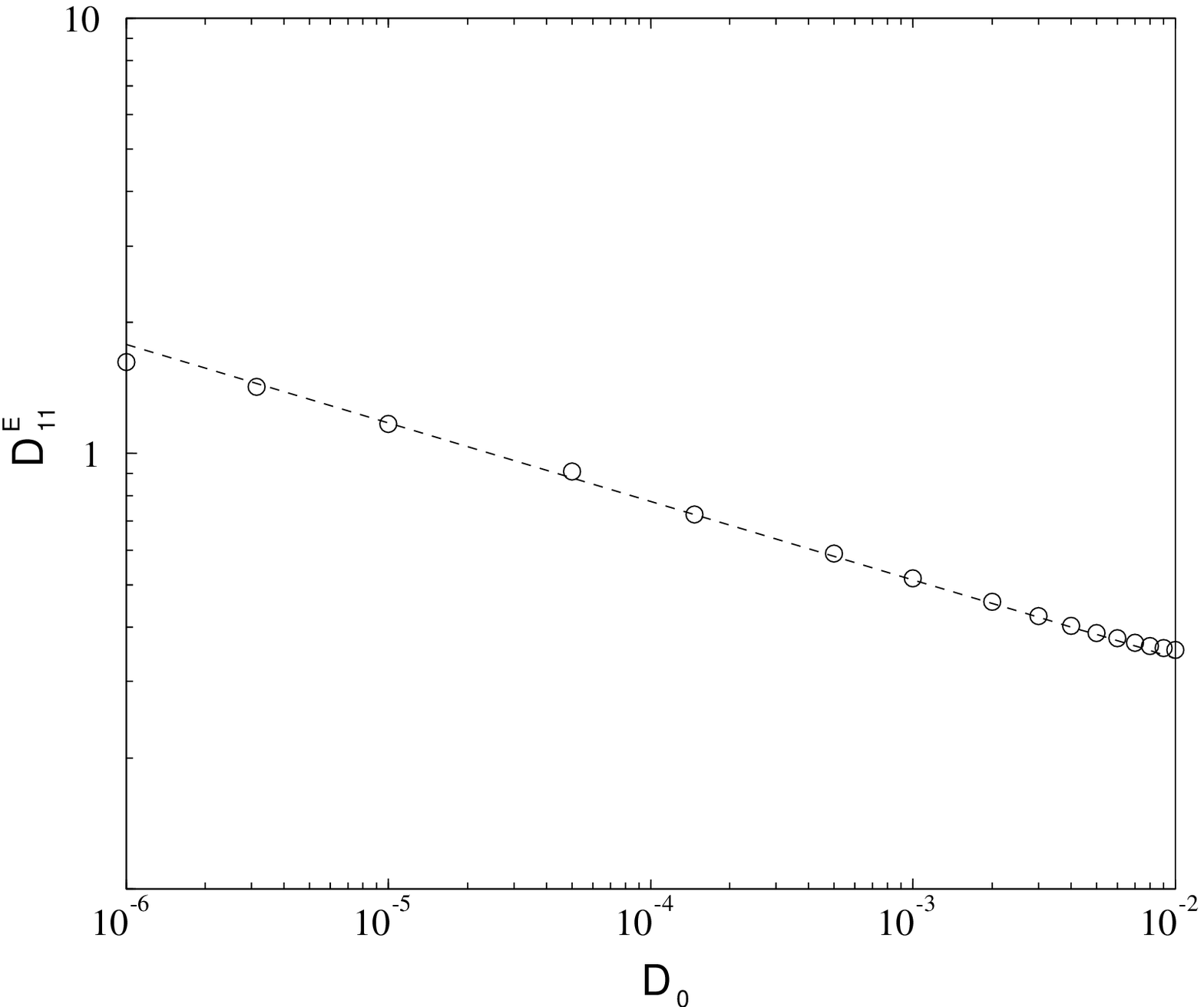,height=7cm,width=7cm}}
\end{center}
\caption{The diffusion coefficient $D^E_{11}$ as a function of $D_0$ for
the frequency 
of the roll oscillation $\omega = 1.1$.
The diffusivities are reported in units of $\psi_0$. 
The best-fit (dashed) line has the slope $-\beta= -0.18$.}
\label{d_vs_d0_anomal}
\end{figure}
%------------------------------------------------
The curve $D^E_{11}$ {\em versus} $D_0$ is
shown in Fig.~\ref{d_vs_d0_anomal}. The data are well fitted by a
straight line with slope $\beta\simeq 0.18$, confirming the presence
of anomalous diffusion at $D_0=0$.

\begin{figure}
\begin{center}
\mbox{\psfig{file=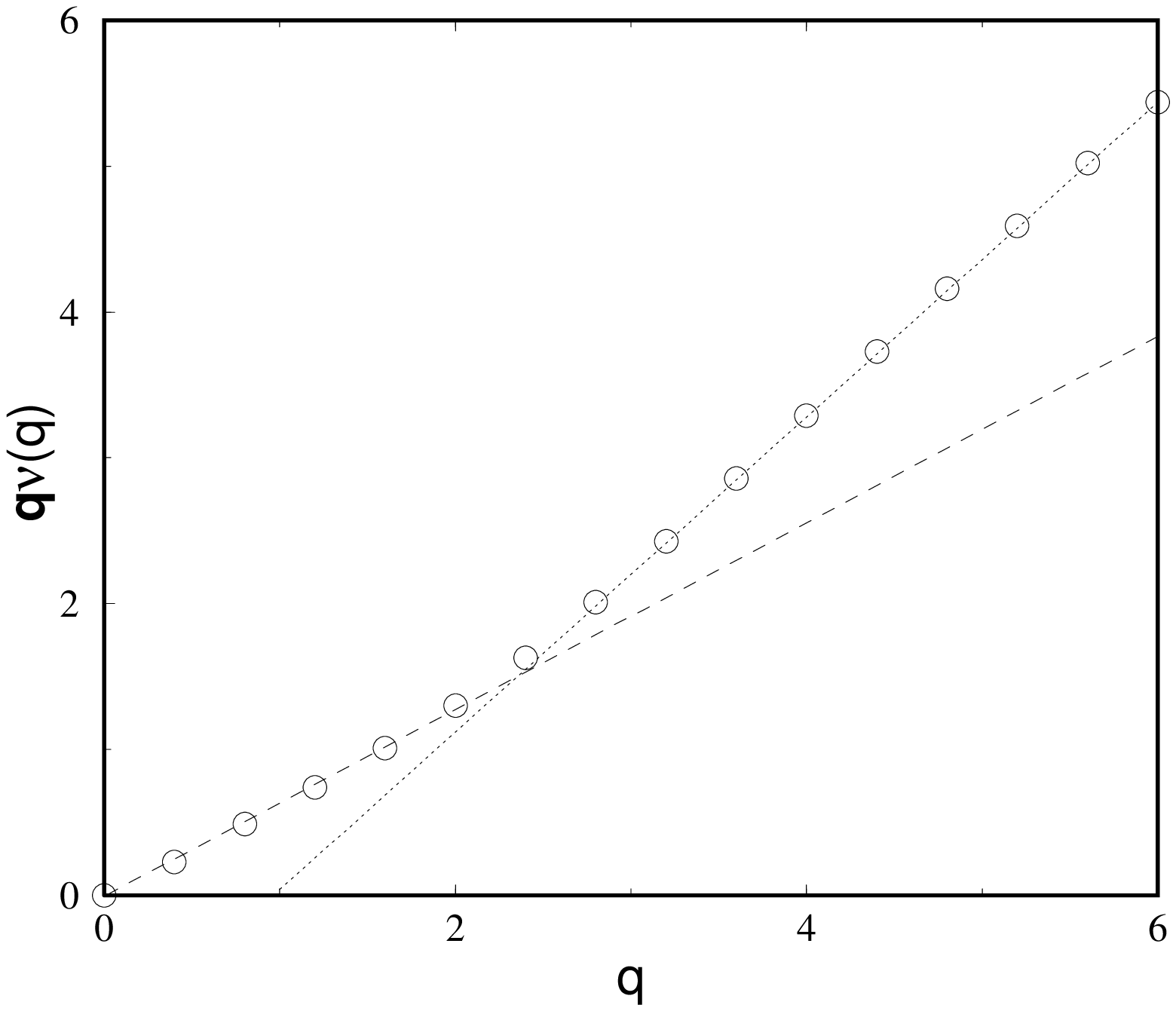,height=7cm,width=7cm}}
\end{center}
\caption{The measured scaling exponents $q\,\nu(q)$'s 
(joined by dot-dashed straight lines)
of the moments of the displacement $\Delta x$,
as a function  of the order $q$. 
The dashed line corresponds to $0.65 \;q$ while the dotted line 
corresponds to $q-1.04$.}
\label{lq_vs_q}
\end{figure}

Fits similar to that one presented in Fig.~\ref{d_vs_t} have been obtained 
also for other values of the order $q$, ranging between $0$ and 
$6$. The results are summarized in Fig.~\ref{lq_vs_q},
%------------------------------------------------
%-------------------------------------------------
where the exponents $q\,\nu(q)$'s 
are shown as a function of $q$. 

Some remarks are in order. The curve, as a whole, displays a nonlinear
behavior, the first clue of anomalous diffusion in the 
strong sense. Two linear regions are actually present:
the first one behaves  up to $q\sim 2$, the second elsewhere.
The two linear 
regions are associated to two different mechanisms in the diffusion
process, as suggested by the following simple considerations. 
For small $q$'s, i.e. for the core of the probability distribution function 
$F(\Delta x,t)$, only one exponent $\nu_1\equiv\nu(q)\simeq 0.65$ for 
$q \lesssim 2$ fully characterizes 
the diffusion process even if anomalous.
The typical, i.e. non rare, events obey a (weak) anomalous diffusion process, 
roughly speaking, one can say that at scale $l$ the characteristic time 
$\tau(l)$ behaves like 
\begin{equation}
\tau(\l) \sim l^{1/\nu_1}.
\label{mazz1}
\end{equation}
 On the other hand, the behavior $q \,\nu(q)\simeq q-1.04 $ suggests that the 
large deviations are essentially associated to ballistic 
transport
\begin{equation}
\tau(\l) \sim l
\label{mazz2}
\end{equation} 
basically due to the mechanism of synchronization
between the circulation in the cells and their global oscillation.

\begin{figure}
\begin{center}
\mbox{\psfig{file=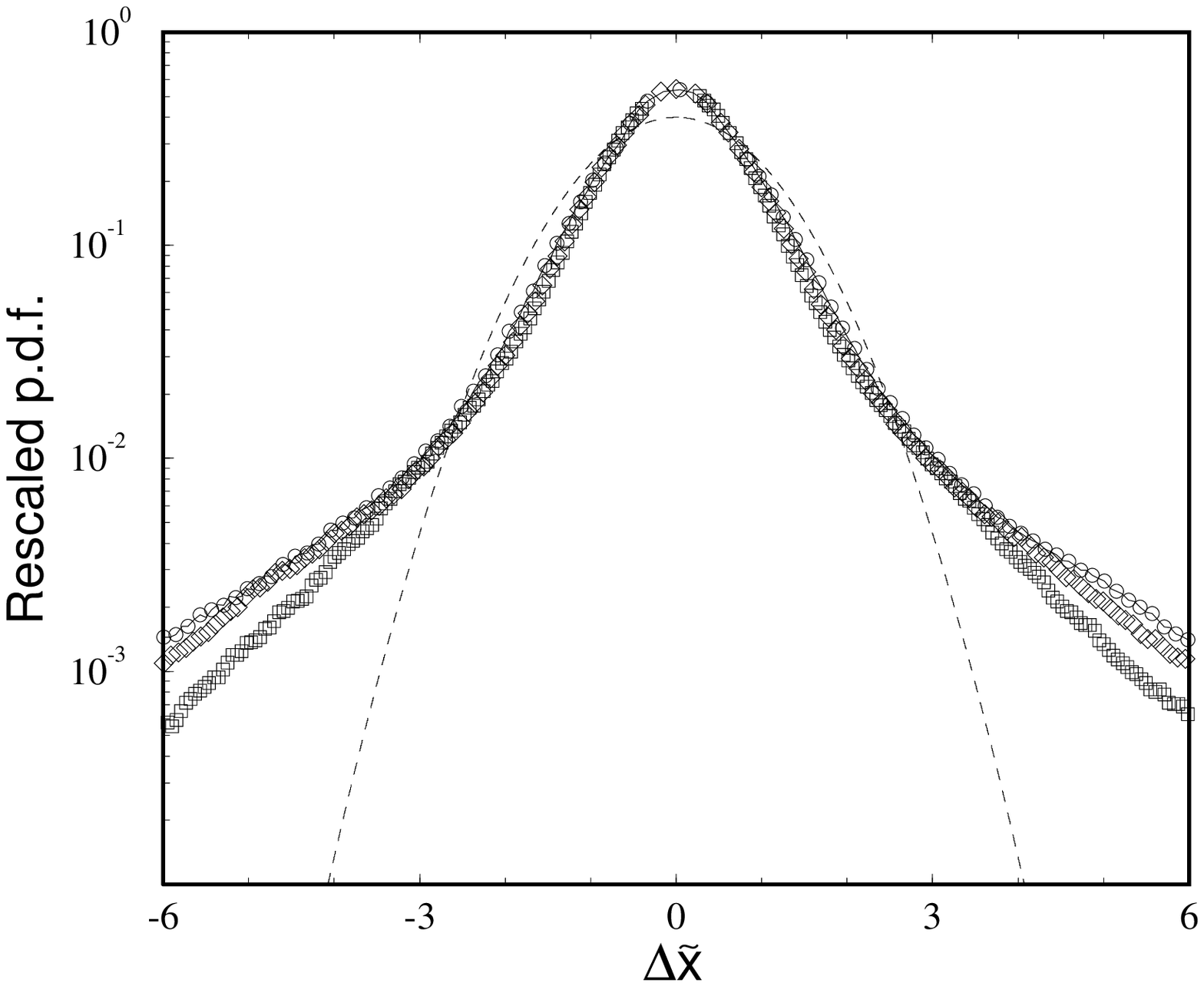,height=7cm,width=7cm}}
\end{center}
\caption{The normalized probability distribution function
$P(\Delta x(t)/\tilde{\sigma})$ vs 
$\Delta \tilde{x} \equiv \Delta x /\tilde{\sigma}$
for the three times 
$t_1=500$ (circles), $t_2=2 t_1$ (diamond) and $t_3=2 t_2$ 
(squares).  The dashed line represents the  
Gaussian function, shown for comparison. 
}
\label{pdf_anom}
\end{figure}

The character of strong anomalous diffusion, already evident from  
Fig.~\ref{lq_vs_q}, is  highlighted by the behavior of the probability 
density function (p.d.f.), 
$P(\Delta x, t)$, of the displacement 
$\Delta x = x(t) - x(0)$ at different times t's. As already pointed out
in Sec.~2.3, weak and strong anomalous diffusion can indeed carefully be 
discriminated by means of the simple scaling law (23), which
is not compatible with the strong anomalous diffusion.
We have considered three different probability distribution functions 
for three different times, namely:
$P_1(\Delta x,t_1)$, $P_2(\Delta x,t_2)$ and $P_3(\Delta x,t_3)$, with
$t_3=2t_2=4t_1$ and $t_1$, $t_2$ and $t_3$ are well inside the scaling
regions of the moments of the displacement.
In Fig.~\ref{pdf_anom} 
%------------------------------------------------
%-------------------------------------------------
the three renormalized probability distribution functions 
$P(\Delta x(t)/\tilde{\sigma})$, with  
$\tilde{\sigma}=\exp\langle \ln \mid \Delta x (t) \mid \rangle$, 
at $t=t_1$, $t_2$ and $t_3$, respectively, are shown.

Some remarks are in order.
The first one concerns the tails of the three probability distribution 
functions which are
clearly  much higher than those of the Gaussian one. 
The strong character of the anomalous diffusion can be easily 
investigated by observing that if the scaling law (23) holds then the 
three functions 
rescaled probability distributions at different times 
have to be superimposed one on the other. 
As one can see from Fig.~\ref{pdf_anom},
this is not the case in our system. 
In fact, one satisfies (23) but only for small deviations, 
that means that only the cores of the probability distribution functions 
obey the above scaling law. 
This fact is not surprising being a direct consequence of 
the behavior shown for the scaling exponents $q\,\nu(q)$'s.  
It is worth noting that the anomalous behavior disappears for a 
very small changes of the parameters, e.g. for $\omega=1.075$ standard 
diffusion takes place.

\subsection{Standard and anomalous diffusion in the standard map}
\label{s:standard}

The standard map is a well-known symplectic map \cite{Ott},
\cite{LiebLich}, given by:
\begin{equation}
\left\{ \begin{array}{l}
J_{t+1}=J_{t}+\frac{A}{2\,\pi} \sin(2\,\pi\,\theta_{t})\\
\theta_{t+1}=\theta_{t}+J_{t+1} \quad \;\;\;\; {\rm mod} \;1. 
\end{array}
\right.
\label{standard}
\end{equation}
A short review of its phenomenology follows. For $A=0$ it
is integrable
\begin{equation}
\left\{ \begin{array}{l} 
J_{t}=J_{0} \quad \forall\, t \\
\theta_{t}=\theta_{0}+t\,J_{0} \quad {\rm mod}\, 1 ;
\end{array}
\right.
\end{equation}
for all rational $J_{0}$ one has periodic motion on  
{\em resonant} tori.
As soon as $A>0$ the resonant tori disintegrate in pairs of elliptic and 
hyperbolic fixed points according to the Poincar\'e-Birkhoff theorem 
\cite{Ott}. 
In the corresponding regions chaotic motions may occur. 
Still for {\em small} values of $A$ the orbits remain trapped in 
invariant sets. This is due to the persistence under 
small perturbation of the nonresonant (KAM) tori where aperiodic 
motions take place and to the topological constraint imposed 
by the dimensionality of the system.
In two dimensions the chaotic orbits generated by the disintegration of 
resonant tori are restricted to lie between two bounding KAM curves. 
Since the latter ones are invariant and the area enclosed by them is 
conserved by symplecticity the chaotic regions are also invariant and 
no diffusion can take place \cite{Ott}

\begin{figure}
\begin{center}
\mbox{\psfig{file=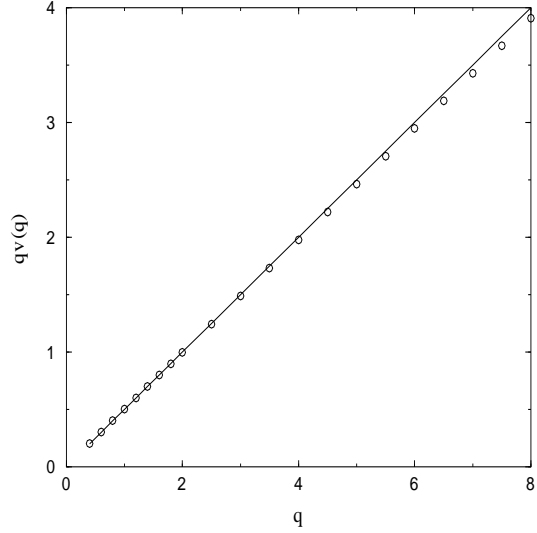,height=7cm,width=7cm}}
\end{center}
\caption{The diffusion exponent $q\,\nu(q)$ as a function of $q$ 
for $A=3.86$. 
The results have been obtained by averaging over $10^{6}$ trajectories 
with $J_{0}=0$.}
\label{exp_norm}
\end{figure}

In order to observe diffusive behaviors one must consider values of 
$A$ large enough so that all KAM tori which connect $\theta=-0.5$ to 
$\theta=0.5$ disappear. 
The critical value of $A$ has been numerically determined to be 
$A_{c}\simeq 0.97$ \cite{Greene}.
For $A>A_{c}$ the phase space can be pictorially described as a chaotic sea 
where islands of regular motions emerge. 
Such islands are related to the existence of accelerator modes i.e. 
stable periodic solutions of 
(\ref{standard})
specified by:
\begin{equation}
\left\{ \begin{array}{l}
J_{t+p}-J_{t}=l \\
A\sum_{t=1}^{p}\sin(2\,\pi \theta_t)= 2 \pi l
\end{array}
\right.
\label{accmod}
\end{equation}
Each accelerator mode can be labelled with the integer pair $(p,l)$.
It is easy to check the stability for $(1,l)$: the symplectic property 
implies that the Jacobian matrix of (\ref{standard}) computed on a stable 
solution must give a pair of complex eigenvalues of magnitude one. 
The condition is satisfied for all $A$ such that:
\begin{equation}
2\,\pi |l|
\leq\, A \,
\leq 2\,\pi |l|\sqrt{1+\left(\frac{2}{\pi\,l}\right)^{2}}
\label{stabilityone}
\end{equation}
Expression (\ref{stabilityone}) implies that period one accelerator
modes with different $l$ cannot coexist. Furthermore the 
length of the interval of $A$ values such that $(1,l)$ is 
stable decreases for large $l$ as:
\begin{equation}
\Delta A \sim \frac{1}{|l|}
\label{da}
\end{equation}
If only one stable island exists no sensitive contribution 
to the diffusion properties of the chaotic orbits is expected.
A trajectory that arrives in the neighbourhood of a stable 
island {\em sticks} to it for a finite amount of time before 
getting away. On average the contribution to the encircling orbit 
is zero. Typically one observes normal diffusion \cite{Hatori}:
\begin{equation}
\langle [J_t-J_0]^{2}\rangle \simeq 2 \,D \,t \;\;\;\;\; t \gg 1 
\label{varnorm}
\end{equation}
where the average is over the initial conditions, the
diffusion coefficient $D$ can be estimated with a crude 
random phase approximation as:
\begin{equation}
D \simeq D_{QL}= \frac{1}{2} \langle \left( J_{t+1}-J_t\right)^2\rangle
=\frac{1}{2}\left( \frac{A}{2\,\pi} \right)^{2} \int_0^1 d\theta \, \sin^2 
(2 \pi \theta) = \frac{A^2}{16 \pi^2}. 
\label{31}
\end{equation}
In Fig.~\ref{exp_norm} the standard diffusion behavior of the exponent 
$q\,\nu(q)$ as a function of $q$ is shown for $A=3.86$ 
i.e. in the absence of any stable islands whereas.
The same standard diffusion occurs for $A=10.492927$ when  
one stable island exists \cite{LiebLich},\cite{Ishi}. 
For both the cases standard diffusion occurs.

The coexistence of many accelerator modes has relevant consequences 
on the diffusive behavior of the system.
For values of $A$ corresponding to such a situation, the variance 
goes asymptotically as  
\begin{equation}
\langle [J_{t}-J_0]^{2} \rangle \sim t^{2 \nu(2)} \;\;\;\;\; t \gg 1\;\;\;\;
{\rm with} \;\;\;\; \nu(2) \neq 1/2.
\label{varanom}
\end{equation}
 
The sticking of the chaotic orbits to stable islands leads to the 
appearance of blocks of long range correlation in the 
sequences of the $J_{t}$ variable. 
This behavior is qualitatively similar to the features of the L\'evy walk.

For the case corresponding to $A=6.9115$, where the fundamental 
accelerator mode $(1,1)$ coexists with the two orbits with period three 
$(3,\pm 3)$,  we have found $2 \nu(2)\simeq1.32$ 
in agreement with previous works \cite{Ishi}.

\begin{figure}
\begin{center}
\mbox{\psfig{file=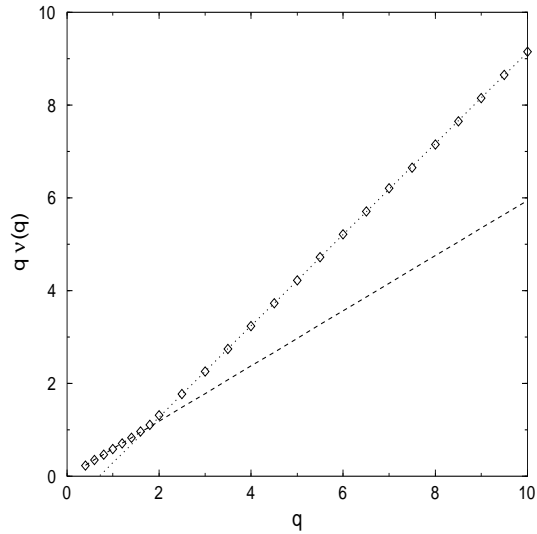,height=7cm,width=7cm}}
\end{center}
\caption{The diffusion exponent $q\,\nu(q)$ as a function of $q$ 
for $A=6.9115$. The averages are over a sample of $10^{6}$ 
trajectories.
The dashed line corresponds to $0.59 \;q$ while the dotted line 
corresponds to $q-0.69$.}
\label{exp_69115}
\end{figure}

On the other hand, in order to investigate the presence of strong
anomalous diffusion, the knowledge of $\nu(2)$ does not characterize
completely the diffusion properties of the orbits.  From the numerical
results (see Fig.~\ref{exp_69115}), one can extrapolate with fair
agreement

\begin{equation}
q \,\nu(q)=\left\{
\begin{array}{cc}
0.59 q       & q \lesssim 2 \\
q-0.69         & q \gtrsim 2.   
\end{array}
\right.
\label{29bis}
\label{stepfun}
\end{equation}

The second moment $\nu(2)\simeq 0.65$ falls in a crossover region between 
the two behaviors.
The feature (\ref{stepfun}) suggests the presence of two mechanisms 
for the typical events and the rare (ballistic) ones as already discussed 
for the $2$-d time dependent flow in the previous subsection.

\begin{figure}
\begin{center}
\mbox{\psfig{file=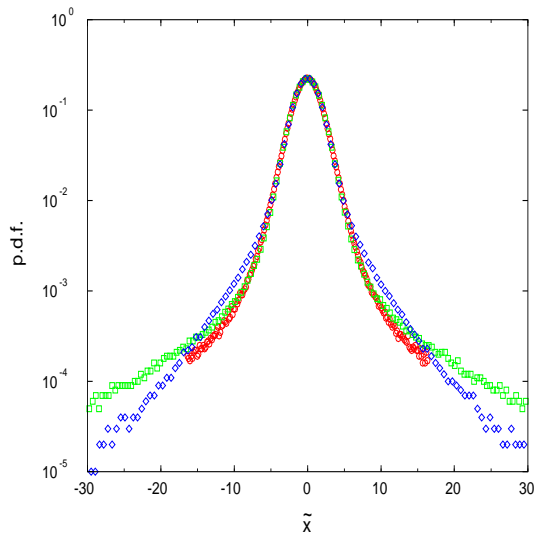,height=7cm,width=7cm}}
\end{center}
\caption{The rescaled probability distribution function 
$P(\Delta J(t)/\tilde{\sigma})$ after $N=4\times 10^6$
iterations, for $A=6.9115$ and $t_1=10^2$ (circles), 
$t_2=10^3$ (squares), $t_4=10^4$ (diamonds) with 
$\tilde{\sigma}(t)=\exp\{ <\ln|\Delta J_t|> \}$.
}
\label{core_69115}
\end{figure}

Such a conjecture can be tested by a direct inspection of the probability 
distribution function. In Fig.~\ref{core_69115} the rescaled  
probability distribution function of 
$\tilde{\Delta J_{t}}={\delta J_{t}}/{\tilde{\sigma}}$ with 
$\tilde{\sigma}=\exp\{<\ln|\Delta J_{t}|>\}$ for different times $t$ is shown.

A further natural question to be addressed is how generic is the strong 
anomalous behavior observed for $A=6.9115$ in the standard map. 
To inquire this point we have studied the diffusion properties 
for $(a),\,\,A=6.8115$, $(b)\,\,A=7.0115$ and 
$(c)\,\,A=6.4717$.
 
%\begin{figure}
%\begin{center}
%\mbox{\psfig{file=exp_abc.eps,height=7cm,width=7cm}}
%\end{center}
%\caption{The diffusion exponent $q \,\nu(q)$ as function of $q$ for 
%$A=6.8115$, $A=7.0115$, $A=6.4717$ compared with the case $A=6.9115$.
%The averages have been performed over $N=10^{6}$ initial conditions 
%with $J_{0}=0$ and up to $T=35282$ iterates.
%}
%\label{exp_abc}
%\end{figure}

The rationale of $(a)$ and $(b)$ is to check the
stability of the result (\ref{stepfun}) with respect to small 
changes of $A$. 
In the case $(a)$ strong anomalous diffusion is observed. 
The behavior of the moments practically coincides with that 
one observed for $A=6.9115$. 
For $(b)\,\,A=7.0115$ we observe normal diffusion: the fact has 
to be related to the destabilisation of the pair $(3,\pm 3)$.
Finally $(c)$ corresponds to the coexistence with $(1,1)$ of 
two other stable islands now generated by the pair of period
five accelerator modes $(5,\pm 5)$. Also in this 
case there is fair evidence of strong anomalous diffusion 
although now the crossover region between the two regimes is 
more extended. 

At the end of this subsection we briefly discuss some results recently 
obtained in \cite{Mappa1,Mappa2} where the transport in $2-d$ symplectic maps 
(practically a modification of the standard map) is considered.
The authors find that the anomalous diffusion is rather rare. This is 
particularly clear in Figs.~ $5$ and $6$ of \cite{Mappa1} where 
the exponent $\nu(2)$ for the kicked Harper map is shown to be always equal
 to $1/2$ apart very narrow regions in the control parameter space.   

\subsection{Anomalous diffusion in $1$-d intermittent maps}

\begin{figure}
\begin{center}
\mbox{\psfig{file=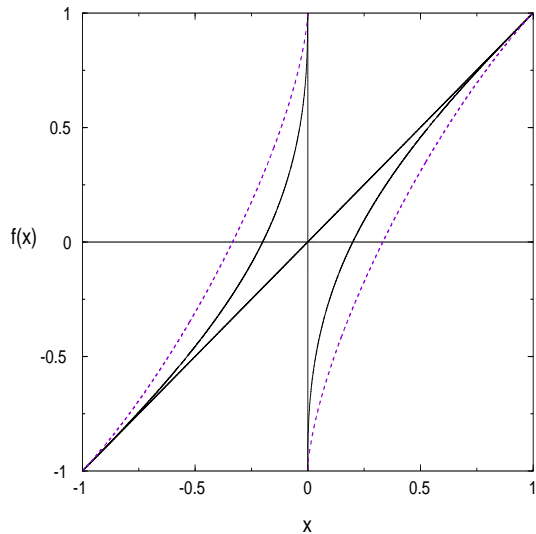,height=7cm,width=7cm}}
\end{center}
\caption{$f(x)$ {\em vs} $x$ from Eq.~(\protect\ref{x1}) 
for $z=2.5$ (full line) and $1.5$ (dashed line). }
\label{mappa}
\end{figure}

Strong anomalous diffusion may also occur in simpler dynamical 
systems of the form
\begin{equation}
\left\{
\begin{array}{ccc}
x_{t+1} & = & f(x_{t}) \\
y_{t+1} & = & y_{t}+x_{t}
\end{array}
\right.
\label{formagenerale}
\end{equation}
where $f(x)$ is a symmetrical Lorenz type map on the interval
$[-1,1]$. Diffusion may then appear in the variable $y_{t}$.
As an example we reconsidered a map previously studied by  
Pikovsky \cite{Pikov}. The function $f(x)$ is  implicitly specified,
$\forall z>1$, for $x \in ]0,1]$ by
\begin{equation}
x=\left\{
\begin{array}{ccc}
\frac{1}{2\,z}[1+f(x)]^{z} &   \hbox{for} &  0<x<\frac{1}{2\,z} \\
f(x)+\frac{1}{2\,z}[1-f(x)]^{z}  &\hbox{for} & \frac{1}{2\,z}<x<1 
\end{array}
\right.
\label{pikovmap}
\end{equation}
with the further requirement of monotonicity. The extension
to the interval $[-1,0[$ is obtained by imposing
\begin{equation}
f(-x)=-f(x)
\label{dispari}
\end{equation}
The resulting shape of $f(x)$ is plotted in Fig.~\ref{mappa}.

Thank to the implicit definition (\ref{pikovmap}),  
by a straightforward computation, it is easy to verify  
that the map has a uniform density invariant distribution. 

As discussed in Sec.~\ref{s:review} anomalous diffusion 
follows from power-law behavior of the correlation function of
the increments of the diffusing variable.
Here the latter property stems from the intermittency of the
dynamics. The mapping (\ref{pikovmap}) induces in the interval
$[-1,1]$ two qualitatively different regions. It is not 
restrictive to focus on $[0,1]$

In the neighbourhood  of $x=1$ the orbits perform laminar motions. 
The basic features of the dynamics are captured \cite{MatdeMar} 
by the polynomial behavior around the unstable fixed point.
\begin{equation}
f(x)=x+\frac{(1-x)^{z}}{2\,z}+o((1-x)^{z})
\label{x1}
\end{equation}
>From (\ref{x1}) it follows that the typical duration of
the laminar phase for an orbit starting from $x \approx 1$ is
$T_{x} \sim (1-x)^{-(z-1)}$. Since the invariant measure is uniform 
a reasonable inference \cite{Pomo} is that the probability 
distribution of having laminar motion of duration $t$ scales as 
\begin{equation}
P_{lam}(t) \sim t^{-\frac{1}{z-1}}.
\label{prob}
\end{equation}

A qualitatively different picture comes from the set 
$[0,\frac{1}{2\,z}]$.
There the mapping has the explicit representation: 
\begin{equation}
f(x)=[(2\,z\,|x|)^{\frac{1}{z}}-1].
\label{esatto}
\end{equation}
In this second region the motion is chaotic and the increments 
$\Delta x_{t}=x_{t+1}-x_{t}$ decorrelate after few steps.
The coexistence of the two regimes prevents the dynamics from having 
a typical timescale so that the increments correlation function 
obeys to a power law   
\begin{equation}
C(\tau)=<\Delta x_{t+\tau}\;\Delta x_{t}> \sim P_{lam}(t) 
\sim t^{-1/(z-1)} 
\label{pkcorr}
\end{equation}
as it has been shown in \cite{Gross}. Anomalous diffusion is
therefore expected for $z>2$.

We studied the diffusion behavior of the dynamics for four 
values of $z$. 

\begin{figure}
\begin{center}
\mbox{\psfig{file=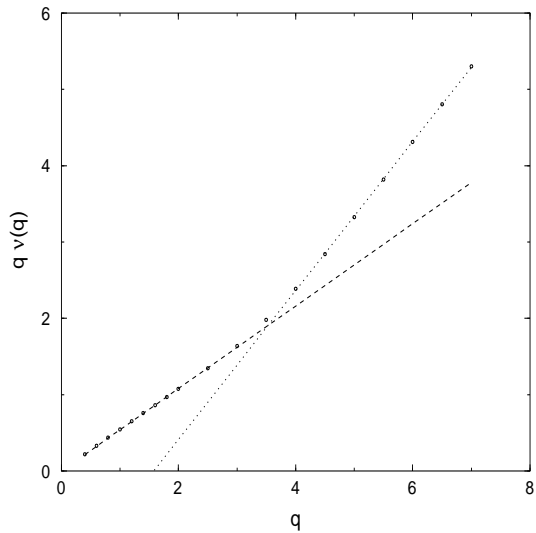,height=7cm,width=7cm}}
\end{center}
\caption{The measured scaling exponents $q\,\nu(q)$'s 
of the moments of the displacement $\Delta x_t$,
as a function  of the order $q$ for $z=1.5$. 
The dashed line $0.5 \;q$ while the dotted line 
corresponds to $q-1.50$.}
\label{exp_p15}
\end{figure}

For $z=1.5$ the integral of the correlation (\ref{pkcorr}) 
is convergent therefore the second moment grows linearly with time. 
We observe standard scaling for all the moments of power $q<3$ 
as shown in Fig.~\ref{exp_p15}.

Let us note that in this case, since $\nu(2)=1/2$, the diffusion seems to be 
standard. Actually, because of the nonlinear behavior of $q \,\nu(q)$, one can 
say that this is an example of strong anomalous diffusion.

\begin{figure}
\begin{center}
\mbox{\psfig{file=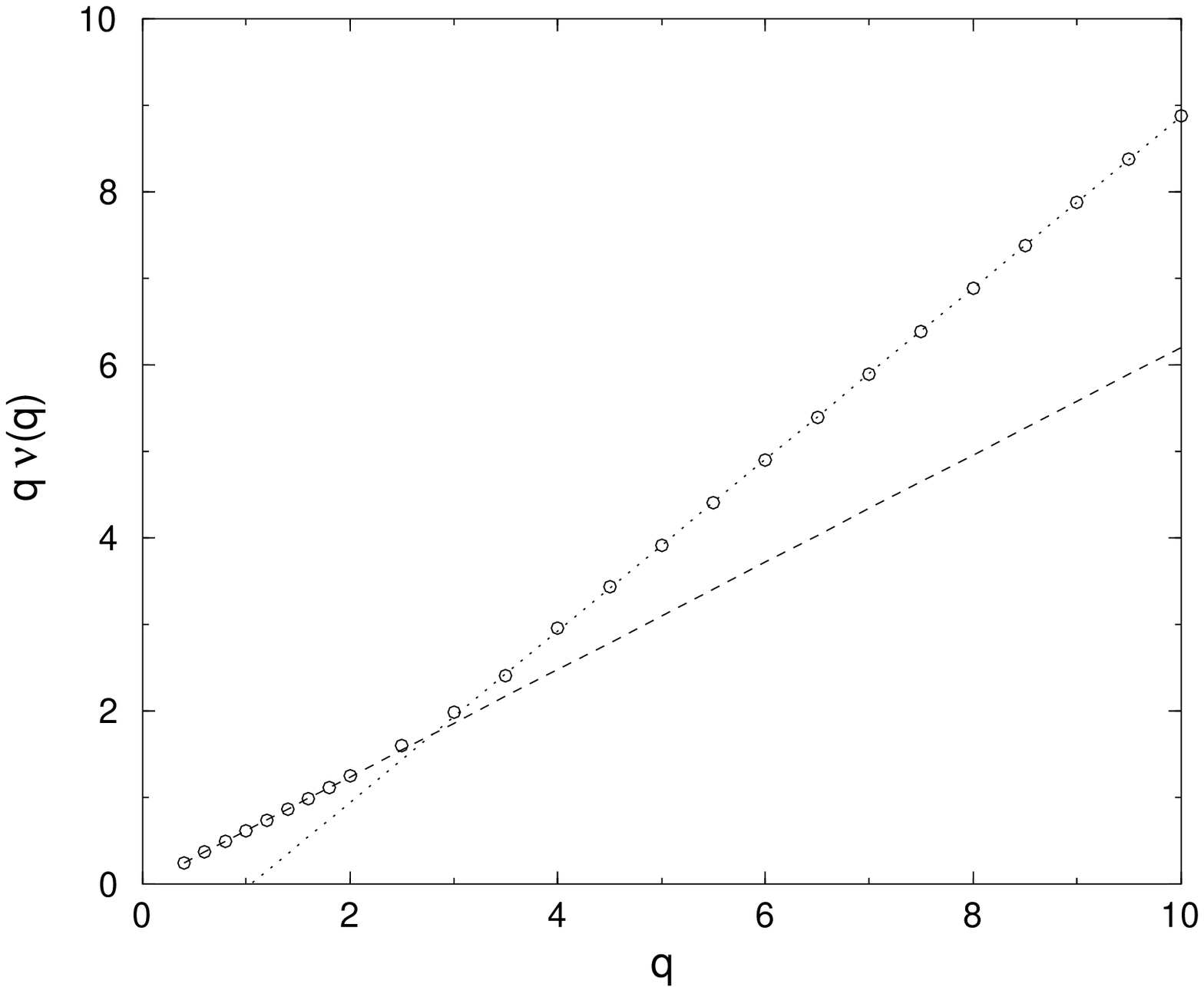,height=7cm,width=7cm}}
\end{center}
\caption{The same as in Fig.~\protect\ref{exp_p15} for $z=2.0$. 
The dashed line corresponds to $0.61 q$ while the dotted line corresponds 
to $q-1.03$.}
\label{exp_p2}
\end{figure}

For $z=2$ the mapping coincides with (\ref{esatto}) in all the 
interval $[0,1]$. This allows the use of a faster iteration 
algorithm.  
The value $z=2$ is also critical for the appearance of anomalous 
diffusion, in the sense $\nu(2)>1/2$, due to the divergence of 
$\sum_{\tau=1}^{\infty} C(\tau)$. 
Actually the plot of the diffusion exponent $q\,\nu(q)$ versus 
$q$ gives a fair evidence for strong 
anomalous diffusion. In Fig.~\ref{exp_p2} the moments larger
than the fourth are in agreement with a ballistic dynamics 
$q \,\nu(q)\simeq q-0.54$ while for moments with $q<2$ it results  
$q \,\nu(q)\simeq 0.61 \, q$.

\begin{figure}
\begin{center}
\mbox{\psfig{file=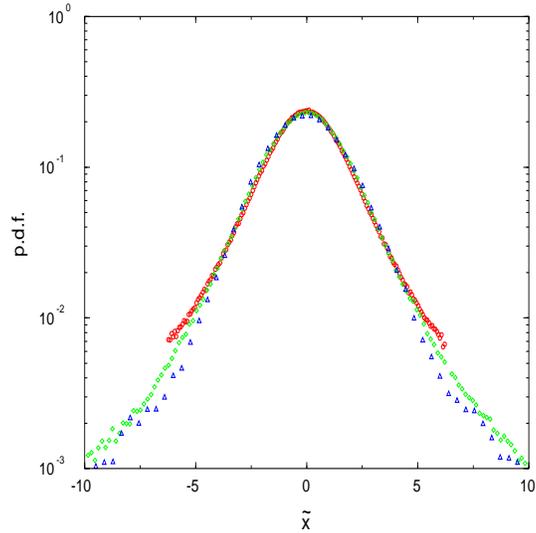,height=7cm,width=7cm}}
\end{center}
\caption{The rescaled probability density as in Fig.~\protect\ref{core_69115} 
but for the Pikovsky map when $z=2$ in Eq.~(\protect\ref{formagenerale}).}
\label{pdf_p2}
\end{figure}

The non Gaussian behavior of the core of the probability distribution 
function appears also from the direct inspection of the distribution as 
shown in Fig.~\ref{pdf_p2} where the probability distributions functions 
calculated for different times and rescaled by $\tilde{\sigma}$ are 
compared with the normal distribution.   

%The investigation of higher values of $z$ doesn't change the 
%qualitative picture. 
%The tails of the PDF for the diffusing 
%variable $y$ are dominated by the laminar r\'egime of the dynamics 
%and give rise to a ballistic law for the moments higher than the second. 
%Lower moments obey to a superdiffusive law with $\nu(q,z)>0.62...$  
%for a given $z$ and monotonously increasing with $z$. The last phenomenon 
%is to be explained with the increase of the space occupied by the laminar 
%region in the interval $[-1,1]$.
 
\begin{figure}
\begin{center}
\mbox{\psfig{file=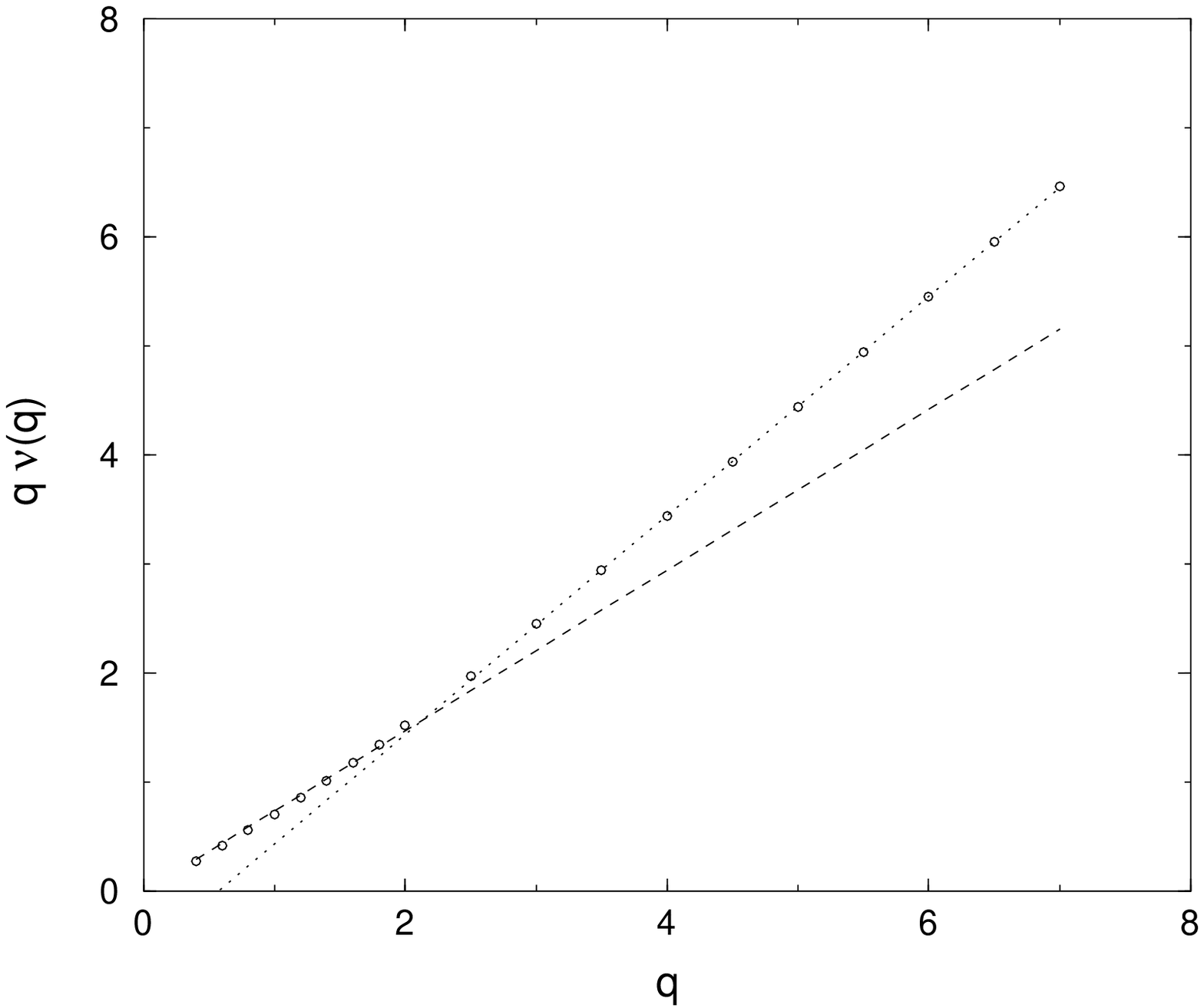,height=7cm,width=7cm}}
\end{center}
\caption{The same as Fig.~\ref{exp_p15} but for $z=2.5$. 
The dashed line corresponds to $0.73 q$ while the dotted line 
corresponds to $q-0.54$.}
\label{exp_p25}
\end{figure}

\begin{figure}
\begin{center}
\mbox{\psfig{file=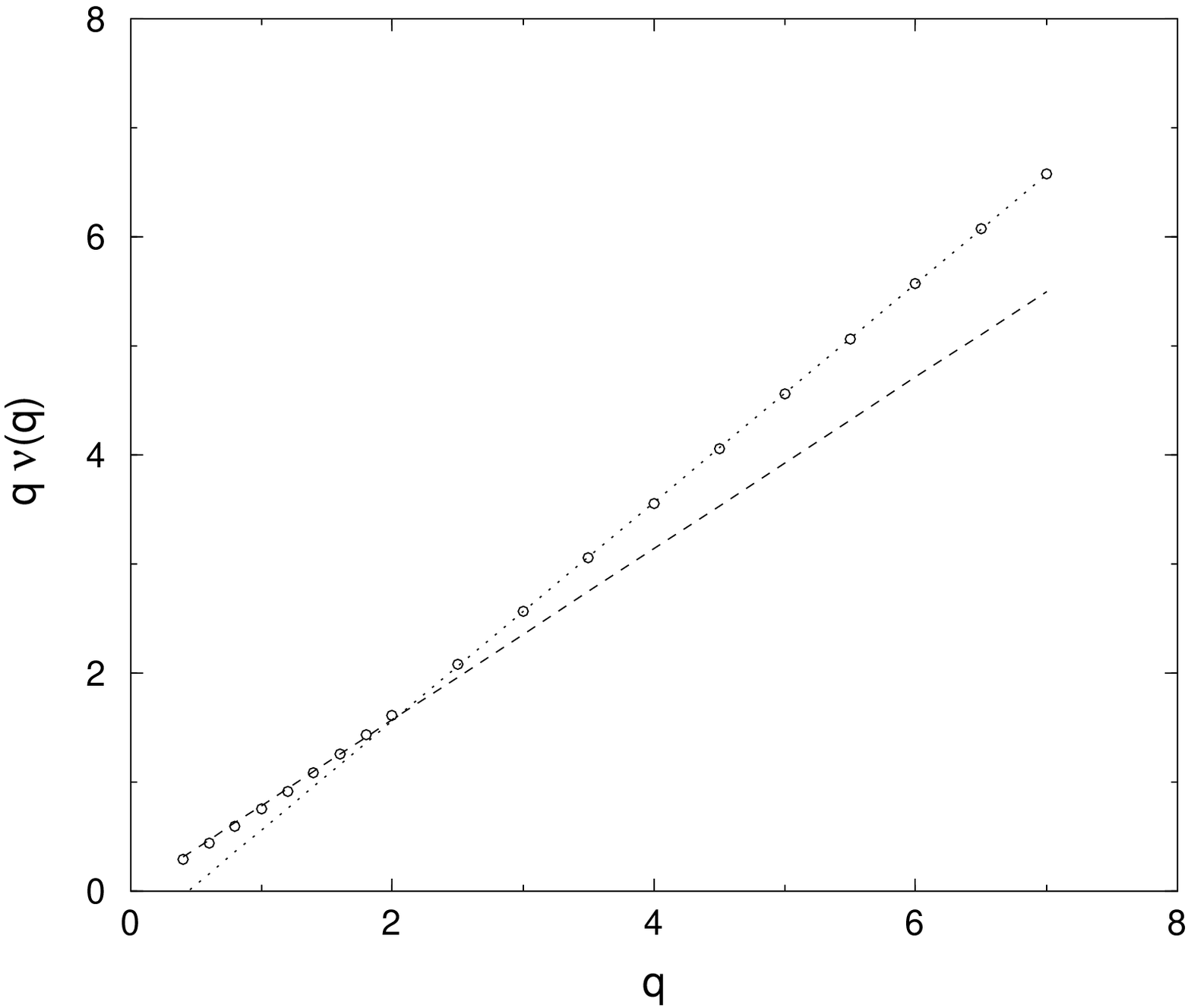,height=7cm,width=7cm}}
\end{center}
\caption{The same as Fig.~\ref{exp_p15} but for $z=3.0$. 
The dashed line corresponds to $0.78 q$ while the dotted line 
corresponds to $q-0.44$.}
\label{exp_p3}
\end{figure}

In Fig.~\ref{exp_p25} and \ref{exp_p3} $q\,\nu(q)$ versus $q$ is reported for 
$z=2.5$ and $z=3.0$. 

As in \cite{Pikov} we can infer from the previous analysis
the following form of the PDF of the diffusing variable $y_{t}$
for $z \ge 2$:
\begin{equation}
P(y,t)=[1-P_{lam}(t)]t^{-\nu_{core}}\,F_{core}(J\,t^{-\nu_{core}})+
P_{lam}(t)[\delta(y-t)+\delta(y+t)].
\label{pikovpdf}
\end{equation}
A consequence of (\ref{pikovpdf}) is 
\begin{equation}
q\,\nu(q)=\left\{
\begin{array}{cc}
q\,\nu_{core}            & q < 2 \\
q- 1/(z-1)               & q > 2   
\end{array}
\right.
\label{29tris}
\label{pikovexp}
\end{equation}
The predictions of (\ref{pikovexp}) are in qualitative agreement with 
the numerical experiments.
For $z<2$ the core of the probability distribution functions becomes 
Gaussian.

\section{Conjectures and remarks}
We have given rather clear numerical evidence of the 
presence of strong anomalous diffusion generated by nontrivial dynamics.
Let us do some remarks and conjectures.
As first we note (Sec.~3) that the anomalous diffusion is, in some sense, 
rare.
For example in the flow mimicking the Rayleigh-B\'enard convection (Sec.~3.1)
the anomalous diffusion occurs only for some precise values of the parameter 
$\omega$. Something similar, even if in a weaker way, happens for the standard
 map (Sec.~3.2) or other $2$-d symplectic maps \cite{Mappa1,Mappa2}.
 Even though for the $1$-d intermittent maps anomalous diffusion could seem a 
generic feature this is not the case. Actually the origin of the 
superdiffusion is in the very peculiar property (marginal instability) of the 
fixed point (see Eq.~(\ref{x1})) destroyed by any generic modifications of the 
map. 
Asymptotic anomalous diffusion, at least that one generated by non 
{\em ad hoc} dynamical systems, seems to be a nongeneric property i.e. it 
disappears as soon as a small perturbation ${\cal} O(\epsilon)$ in the 
evolution law is introduced. However {\em transient} anomalous diffusion 
appears in the perturbed system up to a crossover time 
$t_*\sim \epsilon^{-\alpha}$ while for larger time standard diffusion takes 
place with a rather large diffusion coefficient.
Therefore one can say that, even if asymptotic anomalous diffusion is very 
rare, its ghost is visible \cite{BCVV95}.

In all the cases here considered strong anomalous diffusion has been found 
to correspond to a rather simple shape of the $\nu(q)$:
  
\begin{equation}
q \, \nu(q) \simeq \left\{ \begin{array}{ll}
 \nu_1 \; q   &  q < q_c \\
 q-c  &  q > q_c.
\end{array}
\right.
\label{nu}
\end{equation}
The above behavior suggests the existence of two different mechanisms in the 
diffusion process. The typical (i.e. non rare) events obey a (weak) anomalous 
diffusion process characterized by the exponent $\nu_1$, i.e. the 
characteristic time $\tau(l)$ at the scale $l$ behaves like 
$\tau(l)\sim l^{1/\nu_1}$, but the large deviations are 
essentially associated to a ballistic transport $\tau(l)\sim l$. 
In the flow mimicking the Rayleigh-B\'enard convection the ballistic 
mechanism is due to the synchronization between the circulation in the 
cells and their oscillations. In the standard map it arises from the 
accelerator modes while in the $1$-d intermittent maps is due to the 
marginally unstable fixed point.

We do not claim of course that the above shape (\ref{nu}) 
is universal. It is rather easy, in fact, to build up a toy system for which 
$\nu(q)$ can be any function with the only constraint that $q \, \nu(q)$ 
is a convex function.
Following Elliott et al. \cite{Elliott} we consider the velocity field
\begin{equation}
{\bm u}=(u(y),w)
\label{MH}
\end{equation}   
where $w$ is a constant and $u(y)$ is a given quenched random field.
Neglecting the molecular diffusion the Lagrangian evolution is governed by 
the following equations :
\begin{equation}
\left\{ \begin{array}{l}
\dot{x} = u(y)\\
\dot{y} = w.
\end{array}
\right.
\label{EMH}
\end{equation}
Now, taking  for the sake of simplicity  $w=1$, and $y(0)=0$ 
the following equation holds
\begin{equation}
x(t)-x(0)=\int_0^t dt' \; u(t')
\label{diff}
\end{equation}
which trivially relates the Lagrangian and the Eulerian statistical 
properties of the system.
Thus, from Eq.~(\ref{diff}) it is not difficult understand that adopting a 
suitable multiaffine process \cite{Benzibif,Junei} for $u(y)$ one can find the 
relation defining {\em strong} anomalous diffusion 
\begin{equation}
\langle \mid x(t)-x(0) \mid^q  \rangle \sim t^{q \,\nu(q)} \;\;\;\; 
\nu(q) \neq cost \;\; \; \nu(2) > \frac{1}{2} 
\label{strong}
\end{equation} 
with any proper shape for $\nu(q)$. A fast and efficient algorithm to 
generate a multiaffine process is done in \cite{Bifbof}.

A very interesting open problem is the determination of the 
effective diffusion equation both at long times and large scales 
for a system with strong anomalous diffusion. 
We stress the fact that if $\nu(q) \neq const$ the probability distribution 
$P(\Delta x,t)$ cannot obey a simple generalization of the standard diffusion 
equation. For sure a $P(\Delta x,t)$ corresponding to a $\nu(q)$ such that
(\ref{strong}) holds cannot be obtained from generalized linear diffusion 
equation involving fractional spatial derivative as Eq.~(\ref{225}).
Even other recently proposed linear diffusion equations with fractional 
derivatives both in space and in time are not able to produce such 
nontrivial behavior \cite{Russi}.
        
\section*{Acknowledgments}

We are particularly grateful to M.~Vergassola for very stimulating
discussions and suggestions.  E. Aurell and R. Pasmanter are also
acknowledge for their useful suggestions.  P.M.G. thanks M.H.~Jensen
and M.~Van Hecke for fruitful
discussions and all the CATS staff for the nice
atmosphere at NBI Copenhagen.  P.M.G. is supported by the grant
ERB4001GT962476 from the European Commission.  P.C, is grateful to the
European Science Foundation for the TAO exchange grant.  A.M. was
supported by the ``Henri Poincar\'e'' fellowship (Centre National de
la Recherche Scientifique and Conseil G\'en\'eral des Alpes
Maritimes).  P.C. and A.V. are partially supported by INFM (Progetto
Ricerca Avanzata--TURBO) and by MURST (program no. 9702265437).

\end{document}